\documentclass[aps,prl,preprint,showpacs,groupedaddress]{revtex4}
\usepackage{graphicx}

\begin{document}
\title{The frequency of electron-ion collisions in a hot plasma}

\author{Fedor V.Prigara}
\affiliation{Institute of Microelectronics and Informatics,
Russian Academy of Sciences,\\ 21 Universitetskaya, Yaroslavl
150007, Russia}
\email{fprigara@imras.yar.ru}

\date{\today}

\begin{abstract}
It is shown that the anomalous resistivity, thermal conductivity,
and magnetic pressure of hot plasmas can be explained by the
assumption that the collisional electron-ion cross-section becomes
constant above some critical temperature. This constant is
determined by the size of ion (its electron envelope). It is shown
also that this assumption follows from the consideration of
interaction of a hot plasma with thermal radiation.
\end{abstract}


\maketitle

 It is adopted normally that the only coupling between
ions and electrons in hot plasmas is Coulomb collisions which is
rather weak (see, e.g., [1]). The frequency of electron-ion
collisions in this case is given by the formula [2, 3]

\begin{equation}
\label{eq1}
\nu _{ei} = nv_{Te} \sigma _{ei} \approx 2\pi e^{4}nlog\left( {n\lambda
_{D}^{3}}  \right)/\left( {m_{e}^{2} v_{Te}^{3}}  \right),
\end{equation}

\noindent
where $n = n_{e} = n_{i} $ is the electron and ion density, $v_{Te} $ is the
thermal velocity of electrons, $\sigma _{ei} $ is the electron-ion
collisional cross-section, $m_{e} $and $e$ are the mass and charge of
electron respectively, and $\lambda _{D} $ is the Debye length. It is clear
from equation (\ref{eq1}) that the Coulomb collisional cross-section is very small
at high temperatures, $\sigma _{ei} \propto T_{e}^{ - 2} $, where $T_{e} $
is the electron temperature.

Here we show that this assumption on the electron-ion collisional
cross-section is not correct due to the finite dimensions of ions in a hot
plasma interacting with thermal radiation.

The following effects indicate that the collisional properties of
hot plasmas with $T_{e}$ larger than $T_{0} \cong 3keV$, where
$T_{0} $ is the inversion temperature [8], essentially differ from
those of cold plasmas: 1) anomalous resistivity of a plasma in
strong electric fields [4, 5]; 2) the linear dependence of
magnetic pressure on the magnetic field strength inferred from the
pressure and magnetic field profiles for compact radio sources
[6]; 3) radiation-induced solitary waves in hot accretion disks
[7].

Below we consider these effects in more detail.

Consider current instabilities in a plasma in the absence of magnetic field.
With respect to the voltage-current characteristics, $j = j\left( {E}
\right)$, there are four regimes of electric fields (see also [5]):

1) a classical regime ($j \propto E$) for very low electric fields;

2) quasi-linear regime, $j \cong enc_{s} = const$, where $c_{s}
$is the sound velocity;

3) the heating of a plasma, when the electron temperature, $T_{e} $,
increases, and the voltage-current characteristics is $j \propto E^{\alpha
}$;

4) the regime of the radiative cooling of a plasma, when $T_{e} \approx
const$, and $j \propto E$.

The last regime is normally attributed to the Bunemane instability.

In the region of $j = const$, the collisional electron-ion cross-section is
$\sigma _{ei} \cong \pi \left( {e^{2}/T_{e}}  \right)^{2}$, and the equation
$T_{e} \cong eEl$, where $l = 1/n\sigma _{ei} $ is the mean free path of
electrons, gives $T_{e} \cong \pi ne^{3}/E$. It means the absence of plasma
heating, $T_{e} \approx const$. It is the region of the ion-sound
instability.

Due to the turbulent heating of a plasma, the electron-ion collisional
cross-section becomes a constant, $\sigma _{0} $, which is determined by the
atomic size (see below). In this case, which corresponds to a hot plasma in
a sufficiently strong electric field, $E > E_{c} \cong T_{e} n\sigma _{0}
/e$, the current velocity, $u_{e} $, is proportional to the electric field,
$E$.

In the intermediate region of electric fields, where $u_{e} \cong v_{Te} $,
we have $m_{e} v_{Te}^{2} \cong eE/\left( {n\sigma _{0}}  \right)$, and
therefore $v_{Te} \propto E^{1/2}$. Since

\begin{equation}
\label{eq2}
j = e^{2}E/\left( {m_{e} v_{Te} \sigma _{0}}  \right)
\end{equation}

\noindent
in this regime $j \propto E^{1/2}$. (Compare this result with the relation
$j \propto E^{1/3}$ given by Galeev and Sagdeev [5]).

The critical value of the electric field, $E_{c} $, is determined by
equation

\begin{equation}
\label{eq3} u_{e} = j/en = eE/nm_{e} v_{Te} \sigma _{0} \cong
v_{Te} ,
\end{equation}

\noindent
which gives

\begin{equation}
\label{eq4}
E_{c} \cong m_{e} v_{Te}^{2} n\sigma _{0} /e \cong T_{e} n\sigma _{0} /e.
\end{equation}

It follows from the above considerations, that in a sufficiently hot plasmas
the electron-ion collisional cross-section is a constant, $\sigma _{0} $,
and the frequency of electron-ion collisions is given by a formula

\begin{equation}
\label{eq5}
\nu _{ei} = nv_{Te} \sigma _{ei} \cong nv_{Te} \sigma _{0} .
\end{equation}

The cross-section $\sigma _{0} $ is much larger than the Coulomb collisional
cross-section at these temperatures, and this determines the anomalous
resistivity and thermal conductivity [3] of a plasma.

The origin of the constant cross-section $\sigma _{0} $ is as
follows. A hot plasma with the temperature $T_{e}$ larger than
$T_{0} \cong 3keV$ is intensely interacting with the field of
thermal radiation. At temperatures $T > T_{0} $ the stimulated
radiation processes dominate this interaction [8]. Thermal
radiation induces radiative transitions in the system of electron
and ion which corresponds to the transition of electron from the
free to the bounded state.

Thus, in a hot plasma interacting with thermal radiation, the bounded states
of electrons and ions restore, leading to the change of collisional
properties of a hot plasma. In this case, the electron-ion collisional
cross-section has an order of magnitude of the atomic cross-section, $\sigma
_{0} \cong 10^{ - 15}cm^{2}$.

Consider now a current tube with the cross-section $S = \pi l^{2}$ and the
length \textit{d} placed in a magnetic field \textit{B}. The current tube is
under the action of Ampere's force:

\begin{equation}
\label{eq6}
F = BId/c,
\end{equation}

\noindent
where $I = jS$ is the current, \textit{j} is the current density, and
\textit{c} is the speed of light. It is assumed that the current tube is
perpendicular to the magnetic field force lines.

Suggesting that the quasilinear regime of the voltage-current
characteristics is realized, we can put

\begin{equation}
\label{eq7}
j = enc_{s} ,
\end{equation}

\noindent
where $c_{s} $ is the sound velocity.

The area of the lateral surface of the current tube is $A = \pi ld$, and now
we can calculate the pressure produced by the magnetic field on the current
tube:

\begin{equation}
\label{eq8}
p_{B} = F/A = eBnlc_{s} /c.
\end{equation}

Substituting \textit{l} by the mean free path of electrons with respect to
the scattering on the ions $l_{ei} = 1/n\sigma _{0} $, where $\sigma _{0} $
is the above high-temperature electron-ion collisional cross-section, we
find that

\begin{equation}
\label{eq9}
p_{B} = eBc_{s} /\sigma _{0} c.
\end{equation}

The equation (\ref{eq9}) gives the linear dependence of the magnetic pressure on the
magnetic field strength, contrary to the quadratic dependence given by
magneto-hydrodynamics for cold plasmas (e.g., [4]). For a hot plasma
interacting with thermal radiation, we can assume

\begin{equation}
\label{eq10}
c_{s} = \left( {T_{0} /m_{i}}  \right)^{1/2},
\end{equation}

\noindent
where $T_{0} $ is the inversion temperature, and $m_{i} $ is the mass of
ions.

In this case the magnetic pressure is simply proportional to the magnetic
field strength, the coefficient in this relation being a constant.

Hot plasmas interacting with thermal radiation are characteristic for
accretion disks around compact astrophysical objects such as active galactic
nuclei and young pulsars.

In the case of flat-spectrum active galactic nuclei [8, 9, 10, 11] the
density, temperature, and pressure profiles have a form $n \propto r^{ -
1/2},T \propto r^{ - 1},P = nkT \propto r^{ - 3/2}$. Since for compact radio
sources $B \propto r^{ - 3/2}$, from equation (\ref{eq9}) we obtain $P \propto P_{B}
^{}$. The latter relationship confirms equation (\ref{eq9}). Here the density,
temperature and pressure profiles are the same as those in the
convection-dominated accretion flow (CDAF) models [9]. The magnetic field
profile is different since the expression for the magnetic pressure is
changed. The ratio of magnetic to gas pressure will be fixed similar to the
CDAF models.

It is adopted normally (e.g., [1]) that, in a hot plasma of an accretion
disk, the only coupling between ions and electrons is Coulomb collisions
which is rather weak, and, since the radiation of electrons is much stronger
than that of ions, the ion temperature, $T_{i} $, is much higher than that
of electrons, $T_{e} $.

It follows from the above consideration that such an assumption is not
correct. The coupling between electrons and ions in a hot plasma interacting
with thermal radiation is strong, and therefore such a hot plasma is roughly
the one-temperature plasma, i.e. $T_{e} \approx T_{i} $. This conclusion is
very important for the models of accretion flows in compact astrophysical
objects. In particular, it puts under the question the explanation of the
phenomenon of low-luminosity active galactic nuclei by the models of
accretion flows based on the assumption that $T_{i} $ is much higher than
$T_{e} $ (see, e.g., [1, 9, 10]. It is not surprising that these models
encounter difficulties in the explanation of observational properties of
low-luminosity active galactic nuclei.

Another field where the strong coupling between electrons and ions in a hot
plasma can essentially change the situation is the theory of envelope
(Langmuir) solitary waves [12]. In the last theory the interaction between
Langmuir and ion-sound waves play an important role.

\newpage

\begin{center}
----------------------------------------------------------------------------------------
\end{center}

[1] F.Yuan, and A.Zdziarski, \textit{Luminous hot accretion flows:
the origin of X-ray emission of Seyfert galaxies and black hole
binaries}, Month. Not. R. Astron. Soc. (submitted), E-print
archives, astro-ph/0401058 (2004).

[2] F.F.Chen, \textit{Introduction to Plasma Physics and Controlled Fusion,
Vol. 1: Plasma Physics} (Plenum Press, New York, 1984).

[3] V.P.Silin, Usp. Fiz. Nauk, \textbf{172}, 1021 (2002), Physics-
Uspekhi,\textbf{45}, 955 (2002).

[4] B.B.Kadomtsev, \textit{Collective Phenomena in Plasmas} (Nauka, Moscow,
1988).

[5] A.A.Galeev and R.Z.Sagdeev, in \textit{Handbook of Plasma Physics}, eds.
M.N.Rosenbluth and R.Z.Sagdeev, \textit{Vol. 2: Basic Plasma Physics II},
ed. by A.A.Galeev and R.N.Sudan (North- Holland Physics Publishers,
Amsterdam, 1984).

[6] F.V.Prigara, in \textit{The Magnetized Interstellar Medium}, 8-12
September 2003, Antalya, Turkey, astro-ph/0304243.

[7] F.V.Prigara, \textit{Radiation-induced solitary waves in hot plasmas},
Phys. Plasmas (submitted), physics/0404087.

[8] F.V.Prigara, in \textit{Gamow Memorial International Conference}, 8-14
August 2004, Odessa, Ukraine, astro-ph/0310491.

[9] N.M.Nagar, A.S.Wilson, and H.Falcke, Astrophys. J., \textbf{559}, L87
(2001).

[10] J.S.Ulvestad and L.C.Ho, Astrophys. J., \textbf{562}, L133 (2001).

[11] F.V.Prigara, \textit{Astron. Nachr.}, \textbf{324}, No. S1, 425 (2003).

[12] V.E.Zakharov, in \textit{Handbook of Plasma Physics}, eds.
M.N.Rosenbluth and R.Z.Sagdeev, \textit{Vol. 2: Basic Plasma Physics II},
ed. by A.A.Galeev and R.N.Sudan (North- Holland Physics Publishers,
Amsterdam, 1984).

\end{document}